\documentclass[aps,preprint,preprintnumbers,amssymb,amsfonts,latexsym,groupedaddress,superscriptaddress]{revtex4}

\usepackage{graphicx}
\usepackage{subfigure}
\usepackage{bm}
\usepackage{amsmath}
\usepackage{color}

\newcommand{\bmath}{\begin{mathletters}}
\newcommand{\emath}{\end{mathletters}}
\newcommand{\be}{\begin{eqnarray}}
\newcommand{\ee}{\end{eqnarray}}
\newcommand{\ba}{\begin{array}}
\newcommand{\ea}{\end{array}}

\newcommand{\no}{\nonumber}

\newcommand{\bt}{\beta}

\newcommand{\calG} {\mathcal G}
\newcommand{\calI} {\mathcal I}
\newcommand{\calK} {\mathcal K}
\newcommand{\calL} {\mathcal L}
\newcommand{\calP} {\mathcal P}

\newcommand{\calR} {\mathcal R}

\newcommand{\calU} {\mathcal U}

\newcommand{\calW} {\mathcal W}
\newcommand{\calY} {\mathcal Y}

\newcommand{\rmP} {{\mathrm{P}}}
\newcommand{\rmC} {{\mathrm{C}}}
\newcommand{\Tr} {\mathrm{Tr}}

\newcommand{\tot} {\mathrm{tot}}
\newcommand{\eq}  {\mathrm{eq}}
\newcommand{\rmS} {\mathrm{S}}
\newcommand{\rmB} {\mathrm{B}}
\newcommand{\resum} {\mathrm{resum}}
\newcommand{\exact} {\mathrm{exact}}

\begin{document}
\title{A Continued Fraction Resummation Form of Bath Relaxation Effect in the Spin-Boson Model }
\author{Zhihao Gong}
\affiliation{Physics Department, Zhejiang University, 38 ZheDa Road, Hangzhou, Zhejiang, 310027, China}
\author{Zhoufei Tang}
\affiliation{Physics Department, Zhejiang University, 38 ZheDa Road, Hangzhou, Zhejiang, 310027, China}
\author{Shaul Mukamel}
\affiliation{Department of Chemistry, University of Rochester, Rochester, New York 14627}
\author{Jianshu Cao}
\affiliation{Department of Chemistry, MIT, 77 Massachusetts Ave, Cambridge, MA, 02139, USA}
\author{Jianlan Wu}
\email{jianlanwu@zju.edu.cn}
\affiliation{Physics Department, Zhejiang University, 38 ZheDa Road, Hangzhou, Zhejiang, 310027, China}

\begin{abstract}

In the spin-boson model, a continued fraction form is proposed to systematically resum high-order
quantum kinetic expansion (QKE) rate  kernels, accounting for the bath relaxation effect beyond the
second-order perturbation. In particular, the analytical expression of the sixth-order QKE rate kernel
is derived for resummation. With higher-order correction terms systematically extracted from
higher-order rate kernels, the resummed quantum kinetic expansion (RQKE) approach in
the continued fraction form extends the Pade approximation and can fully
recover the exact quantum dynamics as the expansion order increases.

\end{abstract}

\maketitle

\section{Introduction}
\label{sec1}

In a quantum dynamic process, the interaction between the system and  bath leads to
irreversible energy relaxation and decoherence of the quantum system. The resulting quantum dissipation
can lead to rich quantum phenomena, e.g., quantum phase transition~\cite{sachdev:quantum_phase_transition_book}.
The spin-boson (Caldeira-Leggett) model is a simple but fundamental quantum system, which can be used to interpret
the quantum tunneling and localization in macroscopic systems~\cite{Leggett1981:PhysRevLett,Leggett1987:RMP}.
Gate operations in quantum computation and quantum information are simulated by quantum dissipative dynamics
of multiple spin-boson models, where each qubit is equivalent to an individual spin~\cite{Breuer2002}.
In the study of quantum transport, a fundamental question is to understand the transport
process from a donor to an acceptor in the two-site system~\cite{Nitzan2006}. In the lowest order,
the transfer rate is estimated using Fermi's golden rule (FGR), proportional to the square
of the site-site coupling strength. This second-order transfer rate is
expressed as the Forster theory in energy transfer~\cite{Forster1948:Annphys}
and as the Marcus theory in electron transfer~\cite{Marcus1964:ARPC}.
The non-Markovian relaxation of the surrounding bath can significantly slow down
the transfer process compared to the second-order
prediction~\cite{Zusman1979:CP,Sumi1985:JCP,CaoJung:JCP,Loring1987: JCP,
Mukamel1987:JCP,Mukamel1989:JCP,JLWu2013:JCP,Cao2000:JCP,Laird1990:JCP,Reichman1996:JCP}.
On the other hand, the transfer rate can be optimized
at an intermediate dissipation strength
in a biased two-site system,
which can be further related to the energy transfer optimization in multi-site
systems~\cite{Cao2009:JPCA,JLWu2013:PRL,Plenio2008:NJP,Rebentrost2009:NJP,JLWu2010:NJP,JLWu2012:JCP,Moix2011}.
Within the single excitation manifold,
the two-site system can be viewed as an extension of the spin-boson model, with
possible variations in the boson bath and the bath spatial correlation~\cite{May2004,JLWu2013:JCP}.

As a simple quantum model, the  spin-boson model (or the equivalent two-site system)
 is a benchmark system for the study of quantum dynamic methodologies. In addition to the sophisticated
Feynman-Vernon influence functional~\cite{Feynman1963:AnnPhys}, a straightforward approach of quantum
dissipation is to apply the Nakajima-Zwanzig projection operator~\cite{Nakajima1958,Zwanzig1960:JCP}.
In the lowest second order, we obtain various approximate dynamic equations
from different perturbed terms, e.g., the Redfield equation from the system-bath
interaction~\cite{Redfield1957}, and the FGR rate from the site-site coupling.
The noninteracting-blip approximation (NIBA) extends the FGR rate to a
time-nonlocal description of the detailed time evolution~\cite{Leggett1987:RMP}. To improve the
NIBA prediction, the variational polaron method is a modified second-order perturbation
where the perturbed term is self-consistently determined from equilibrium distribution~\cite{Harris1984:JCP, Wang2014}.
The variational polaron method is more reliable in the unbiased
two-site system with a relatively fast bath.
A more systematic approach beyond the
second-order perturbation is to include higher-order corrections of perturbed terms,
as in the quantum kinetic expansion (QKE) approach~\cite{Mukamel1987:JCP,Mukamel1989:JCP,Laird1990:JCP,
Cao2000:JCP,JLWu2013:JCP,Reichman1996:JCP}.
 In our recent paper~\cite{JLWu2013:JCP}, the higher-order QKE of the site-site coupling
is obtained using an indirect projection operator
technique for a general multi-site system. In the two-site system, all the higher-order
QKE corrections arise from the bath relaxation effect,
whereas in the multi-site system, the higher-order QKE corrections also include quantum interference effects.

A key theoretical concern in the QKE approach is the resummation technique of higher-order rate kernels,
due to two essential reasons. The analytical and numerical difficulties quickly increase as the
expansion order increases. More crucially, the QKE rate kernels can converge slowly and become
divergent as the site-site coupling increases. An appropriate resummation technique can
partially include corrections of all the orders using one or a few higher-order QKE rate
kernels, and avoid the divergence of large site-site couplings.
For the lowest-order
correction, two typical resummation techniques are the
Pade approximation~\cite{Mukamel1987:JCP,Mukamel1989:JCP}
and the Landau-Zener approximation~\cite{Sumi1980:JPSP}. With a factor of 2 difference,
the Landau-Zener approximation is not reliable in the strong dissipation limit,
compared to the Pade approximation.
In a recent paper~\cite{Troy2014:JCP}, a modified resummation approach is
proposed with an optimization according to the equilibrium distribution.
However, any resummation techniques in the lowest order cannot fully account for the extra
knowledge of higher-order QKE rate kernels, and its prediction deviates significantly
from the exact quantum dynamics at some point.

Therefore, a more general resummation technique is
required to systematically include corrections from higher-order rate kernels.
In Ref.~\cite{Cho1996:JCP}, a generalized Pade approximation is developed, which is complicated
in its mathematical formulation and practical application.
Instead, we will extend the physical factorization scheme
in the Pade approximation to the higher-order QKE rate kernels and obtain a simple
continued fraction form, which leads to a systematic resummed quantum kinetic
expansion (RQKE) method. In Section~\ref{sec2}, the derivation of the QKE approach in
the two-site system (the spin-boson model) is briefly reviewed. The time-integrated QKE rates of the first
three orders are numerically computed in a quantum Debye bath.
In Section~\ref{sec3},  the continued fraction resummation form is developed,
and the RQKE rates are numerically compared with the exact results of
both unbiased and biased systems. In this paper, all the exact quantities
are obtained using the hierarchy equation method
~\cite{Tanimura1989:JPSJ,Shao2004:CPL,Yan2005:JCP,Moix2013:JCP}.
In Section~\ref{sec4}, the RQKE
rate kernels are used to predict the detailed population evolution, and are calibrated with
the exact result. In Section~\ref{sec5}, the temperature-dependent
equilibrium population is calculated using the RQKE rates, which is also
compared with the exact stochastic path integral result~\cite{CKLee2012:JCP,Jeremy2012:PRB}.
In Section~\ref{sec6}, we summarize our studies.

\section{Quantum Kinetic Expansion in a Two-Site System}
\label{sec2}

In this section, we briefly review the quantum kinetic expansion (QKE) approach in Ref.~\cite{JLWu2013:JCP}. With respect to
the single excitation manifold, the bare Hamiltonian of a multi-site system is given by
$H_\rmS = \sum_{n}\varepsilon_n |n\rangle\langle n|+\sum_{n\neq m}J_{mn}|m\rangle\langle n|$, where
$|n\rangle$ represents a single-excitation quantum state localized at site $n$,
$\varepsilon_n$ is the excitation energy of site $n$,
and $J_{mn}$ is the site-site coupling strength between sites $m$ and $n$. The bare Hamiltonian of
the surrounding environment is given by $H_\rmB$. The system-bath interaction $H_{\rmS\rmB}$ is considered
to be localized at each site $n$, $H_{\rmS\rmB} = \sum_n H_{\rmS\rmB; n} |n\rangle\langle n|$.
In the site basis representation $\{|n\rangle\}$, the total Hamiltonian is written as
\be
H_{\tot} = \sum_{n}H_n |n\rangle\langle n|+\sum_{nm(n\neq m)}J_{mn}|m\rangle\langle n|,
\label{eq_001}
\ee
with $H_n = \varepsilon_n + H_\rmB + H_{\rmS\rmB;n}$. Here the simplest two-site system coupled
with a harmonic bath can be mapped to the standard spin-boson model.
The time evolution of the total density
matrix $\rho_{\tot}(t)$ follows the Liouville equation,
$\partial_t\rho_{\tot}(t)=-i\calL_{\tot}\rho_{\tot}(t)$,
with $\calL_{\tot}=[H_{\tot}, \cdots]$.
Throughout this paper, the reduced Planck constant $\hbar$ is treated as a unit.
Following  the separation of population and coherence components, the total
Liouville superoperator is formally expressed as a  block matrix,
\be
\calL_{\tot} = \left(\begin{array}{cc}  \mathcal{L}_{\rmP} & \mathcal{L}_{\rmP\rmC}\\
                                        \mathcal{L}_{\rmC\rmP} & \mathcal{L}_{\rmC}
              \end{array}\right),
\label{eq_002}
\ee
where the subscripts $\rmP$ and $\rmC$ denote system population and coherence, respectively.
In the two-site system, the diagonal part of $\calL_{\tot}$ is fully dependent on the
diagonal Hamiltonian elements $H_n$, while the off-diagonal part of $\calL_{\tot}$ arises
from the site-site coupling $J$. Subsequently, we define the partial time propagation
superoperators, $\calU_\rmP(t) = \exp(-i\calL_{\rmP} t)$ and $\calU_\rmC(t) =
\exp(-i\calL_{\rmC} t)$, which can be interpreted as Green's functions in the Liouville space.

\begin{figure}
\includegraphics[width=0.65\columnwidth]{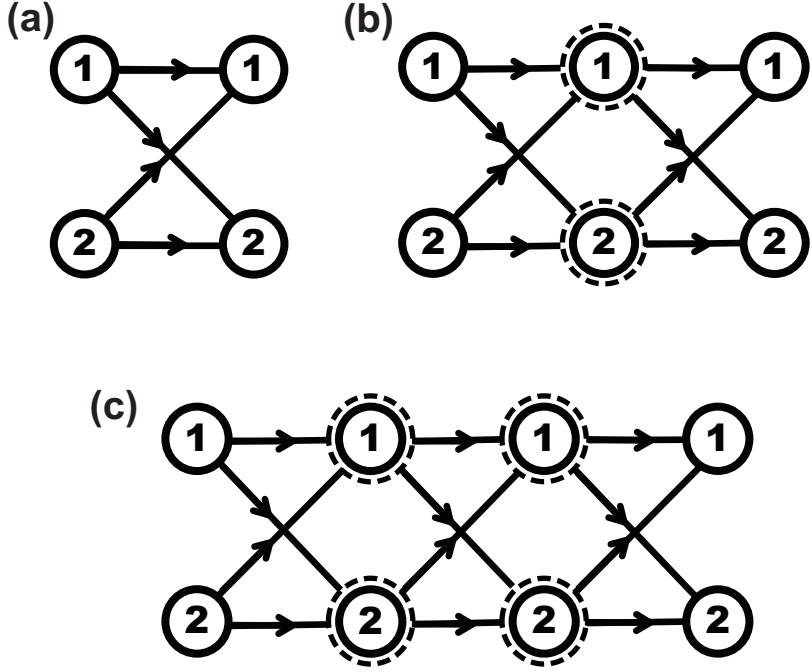}
\caption{The Feynman diagrams of the second- (a), fourth- (b), and sixth-order (c)
quantum rate kernels in the  two-site system (the spin-boson model). The explicit interpretation of each symbol
is provided in text.  }
\label{fig_001}
\end{figure}

An indirect projection operator approach is applied in Ref.~\cite{JLWu2013:JCP} to derive the closed
time evolution equation of the reduced system population
$P(t)$. The initial condition is required to
be a local equilibrium state, $\rho_{\tot}(0) =\sum_n p_n \rho^{\eq}_{\rmB; n} |n\rangle\langle n|$,
where $p_n$ is the initial population of site $n$, and $\rho^{\eq}_{\rmB; n} \propto\exp(-\bt H_n )$
 is the local Boltzman density of bath.
The final time evolution equation of $P(t)$ follows a time-nonlocal convolution form,
\be
\dot{P}(t) = -\int_0^t d\tau \calK(t-\tau) P(\tau).
\label{eq_003}
\ee
The rate kernel $\calK(t)$ is derived as an expansion form of the site-site
coupling $J$, given by $\calK=\calK^{(2)}+\calK^{(3)}+\cdots$.
In the two-site system, all the odd-order terms vanish, and only the even-order terms remain.
Here we introduce a local equilibrium population state matrix,
\be
\rho^{(0)}_{\eq} = \left(\ba{cc} \rho^{\eq}_{\rmB; 1} & 0 \\
                                 0                &\rho^{\eq}_{\rmB; 2} \ea \right),
\label{eq_005}
\ee
and its projection matrix, $\calP^{(0)}_{\eq} = \rho^{(0)}_{\eq} \}\Tr_\rmB\{$,
where $\Tr_\rmB\{\cdots\}$ is the partial trace over bath degrees of freedom.
 The $2k$-th QKE rate kernel
is explicitly given by
\be
&& \calK^{(2k)}(\tau_2, \tau_3, \cdots, \tau_{2k}) \no \\
&=& -(-1)^k\Tr_\rmB\{[\calR(\tau_{2k})\delta\calU_\rmP(\tau_{2k-1})] \no \\
&&~~~ [\calR(\tau_{2k-2})\delta\calU_\rmP(\tau_{2k-3})]\cdots  \calR(\tau_{2})\rho^{(0)}_{\eq} \},
\label{eq_006}
\ee
where
$\delta\calU_\rmP(t)=\calU_\rmP(t)-\calP^{(0)}_{\eq}$ is the pure dissipative propagation
superoperator, vanishing in Markovian dynamics, and
$\calR(t)=\calL_{\rmP\rmC}\calU_\rmC(t)\calL_{\rmC\rmP}$ is the population-to-population transition
superoperator.  Thus, high-order ($k\ge2$) QKE rate kernels reflect dynamics of population fluctuation around
the local equilibrium state due to the bath relaxation effect of $\delta\calU_\rmP(t)$.
Equation~(\ref{eq_006}) is equivalent to the previous expression of Eq.~(15) in Ref.~\cite{JLWu2013:JCP},
but in a more concise form.
The Feynman diagram technique is applied to visualize these quantum rate kernels in Fig.~\ref{fig_001},
which is also simplified in notation compared to previous diagrams in Ref.~\cite{JLWu2013:JCP}. In detail,
each initial and final numbered circle represents a local equilibrium population state,
$\rho^{\eq}_{\rmB; n}|n\rangle\langle n|$, at the corresponding site
$n(=1, 2)$.
Each intermediate dashed circle represents the dissipative propagation $[\delta\calU_\rmP(t)]_n$
of a system-bath entangled population state,
$\rho_{\rmP; n}(t)=[\rho_{\tot}(t)]_{nn}$. Unlike the notation in Ref.~\cite{JLWu2013:JCP},
each arrowed line represents a population-to-population transition $[\calR(t)]_{mn}$,
as a density flow from population to coherence and back to population,
$[\calR(t)]_{mn} = |J_{mn}|^2 [\calU_{\rmC; mn}(t)+\calU_{\rmC; nm}(t)]$.

The formal expression of $\calK^{(2k)}(\tau_2, \tau_3, \cdots, \tau_{2k})$ in Eq.~(\ref{eq_006})
is derived for an arbitrary environment, beyond the spin-boson model.
Next we assume that the bath is harmonic and $H_{\rmS\rmB}$
follows a bilinear form. With the creation ($a_i^{+}$) and annihilation ($a_i$)
operators for the $i$th harmonic oscillator, the bath-coupled Hamiltonian at local site $n$
reads
\be
H_n = \varepsilon_n + \sum_i \omega_i a_i^{+}a_i + \sum_i \omega_i x_{ni}\left(a_i^{+} + a_i \right),
\label{eq_007}
\ee
where the coefficient $x_{ni}$ denotes the system-bath coupling strength reduced by the
frequency $\omega_i$ of the $i$th harmonic oscillator.
The QKE rate kernels in Eq.~(\ref{eq_006}) are transformed into the time correlation functions of
the displacement operator, $G_n = \mathrm{exp}\left[\sum_i x_{ni}(a_i^{+} - a_i)\right]$,
which can be obtained by the cumulant expansion. If the bath coupling is identical at
each system site, the explicit expression of the second-order rate kernel reads
\be
\mathcal{K}_{mn(\neq m)}^{(2)}(\tau_2)
&=& - 2 |J_{mn}|^2\mathrm{Re}~\exp\{-[i\tilde{\varepsilon}_{mn}\tau_2+s_{mn}g(\tau_2)]\},
\label{eq_009}
\ee
where $\tilde{\varepsilon}_{nm}=\tilde{\varepsilon}_{n}-\tilde{\varepsilon}_{m}$ is the modified
site excitation energy detuning with $\tilde{\varepsilon}_n =\varepsilon_n - \sum_i \omega_i x^2_{ni}$,
and the coefficient $s_{mn}$ arises from the site-site `spatial' correlation.
For the standard spin-boson model, a perfectly negative correlation leads to $s_{mn(\neq m)}=4$,
while for the regular energy transfer system, a $\delta$-spatial correlation leads to $s_{mn(\neq m)}=2$.
Thus, the two-site system under the $\delta$-spatial correlation is equivalent to
the spin-boson model with a doubled dissipation strength (reorganization energy).
The time correlation function of the displacement operator excluding the spatial dependence is
\be
g(t) &=& \int_0^{\infty}\mathrm{d}\omega [J(\omega)/\omega^2][(1-\cos\omega t)\coth(\beta\omega/2) \no \\
&&~~~~+ i \sin\omega t],
\label{eq_010}
\ee
where $J(\omega)=\sum_i \omega^2_i x^2_i \delta(\omega-\omega_i)$ is the bath spectral density.
Equation~(\ref{eq_009}) is the same as the rate kernel in the NIBA approach~\cite{Leggett1987:RMP},
and its time integration recovers the FGR rate.
In Ref.~\cite{JLWu2013:JCP}, the fourth-order QKE rate kernel is derived for a general multi-site system.
The simplified expression of $\calK^{(4)}(\tau_2, \tau_3, \tau_4)$
for the two-site system with the $\delta$-spatial correlation
is provided in Appendix~\ref{appa}. Furthermore, we extend to the sixth-order QKE
rate kernel, and the explicit expression of 16 terms is also shown in Appendix~\ref{appa}.

\begin{figure}
\includegraphics[width=0.75\columnwidth]{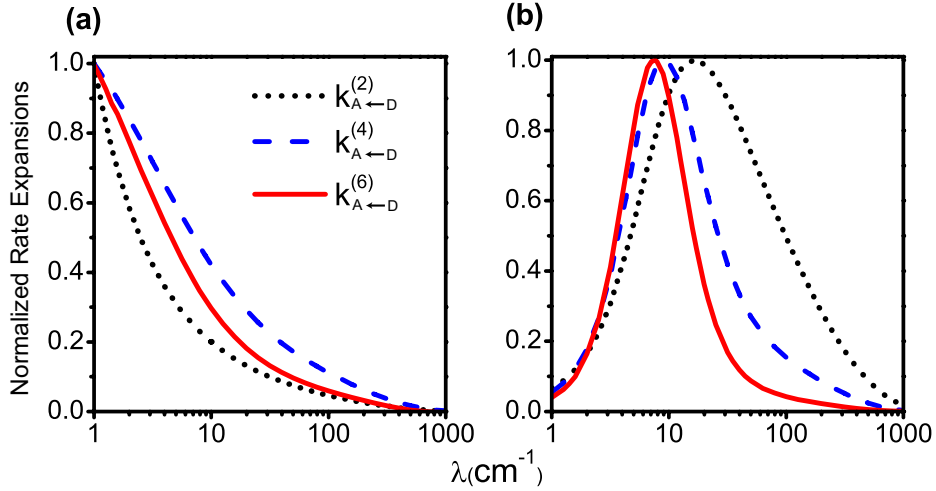}
\caption{The normalized time-integrated forward transfer rate expansions of the first three orders (a) in the unbiased
system with $\varepsilon_{12}=0$, and (b) in the biased system with $\varepsilon_{12}=100$ cm$^{-1}$.
The Deybe frequency of the coupled bath is $\omega^{-1}_D = 100$ fs, and the temperature is $T=300$ K.
The normalization is realized by (a) $k^{(2k)}_{A\leftarrow D}(\lambda)/k^{(2k)}_{A\leftarrow D}(\lambda=1 \mathrm{cm}^{-1})$,
and (b) $k^{(2k)}_{A\leftarrow D}(\lambda)/k^{(2k)}_{A\leftarrow D; \mathrm{max}}$.
In each figure, the dotted black line is the second-order result $k^{(2)}_{A\leftarrow D}$, the dashed blue line
is the fourth-order result $k^{(4)}_{A\leftarrow D}$, and the solid red lines is the sixth-order result $k^{(6)}_{A\leftarrow D}$. }
\label{fig_002}
\end{figure}

Before investigating the resummation technique in next section, we numerically calculate the quantum
rate kernels of the first three orders. Both unbiased and biased two-site systems are considered with
$\varepsilon_{12}=0$ and $100$ cm$^{-1}$. To be compared with the calculation of the
hierarchy equation~\cite{Tanimura1989:JPSJ,Shao2004:CPL,Yan2005:JCP,Moix2013:JCP},
 a quantum bath with the Debye spectral density is applied, given by
\be
J(\omega) = \Theta(\omega) \left(\frac{2\lambda}{\pi}\right) \frac{\omega\omega_D}{\omega^2+\omega_D^2},
\label{eq_011}
\ee
where $\Theta(\omega)$ is the Heaviside step function of $\omega$, $\lambda$ is the reorganization energy,
and $\omega_D$ is the Debye frequency. For simplicity, we introduce the high-temperature approximation,
leading to
\be
g(t) \approx \frac{2\lambda}{\beta\omega_D}\left[|t|-\frac{1-e^{-\omega_D|t|}}{\omega_D}\right]
+i \mathrm{Sign}(t)\lambda\frac{1-e^{-\omega_D|t|}}{\omega_D},
\label{eq_012}
\ee
where $\mathrm{Sign}(t)$ is the sign function of $t$. In our calculation, the Debye frequency is
$\omega^{-1}_D = 100$ fs, and the temperature is $T=300$ K. We focus
on the time-integration of rate kernels, $\calK^{(2k)}=\int_0^\infty d\tau_2 \cdots \int_0^\infty d\tau_{2k}
\calK^{(2k)}(\tau_2, \cdots, \tau_{2k})$, which can be viewed as the time-integrated effective rate matrix,
especially for over-damped dynamics. Since the $2k$-th rate kernel is proportional to the $2k$-th power
of the site-site coupling $J$, we normalize effective rates to remove the $J$-dependence.
The normalization is over the maximum value $\calK^{(2k)}_{\mathrm{max}}$ for the biased system,
and over the value of the minimum reorganization energy ($\lambda=1$ cm$^{-1}$)
for the unbiased system. Due to the heavy computational duty in a multi-time
integration, the Monte Carlo simulation of $10^{12}$ samples is applied to the calculation of
$\calK^{(6)}$ for convergence. Figure~\ref{fig_002} presents the numerical results of the forward transfer rate
expansions $k^{(2k)}_{A\leftarrow D}$ from the donor site 1 to the acceptor site 2,
which will be used for the resummation technique in next section. We find that
$k^{(2k)}_{A\leftarrow D}$ monotonically decreases with the reorganization energy $\lambda$ in the
unbiased system, whereas $k^{(2k)}_{A\leftarrow D}$ is maximized in an intermediate value of $\lambda$ in the
biased system.

\section{Resummation of QKE Rate Kernels in a Continued Fraction Form}
\label{sec3}

In the previous section, we present the explicit expansion forms of rate kernels in
the two-site system (the spin-boson model) using the QKE approach.
For a small site-site coupling strength, the full quantum kinetic
rate kernel can be obtained straightforwardly as the sum of $\calK^{(2k)}$ up to
a converged expansion order. For a large site-site coupling strength, this simple summation
cannot be applied since $\calK^{(2k)}$ diverges as the expansion order increases. Instead,
a resummation technique is required for a converged result, with one or more high-order
corrections of $\calK^{(2k)}$ $(k\ge 2)$. For the leading-order QKE correction $\calK^{(4)}$,
various resummation methods, e.g., the Pade approximation~\cite{Mukamel1987:JCP} and the Landau-Zener
approximation~\cite{Sumi1980:JPSP}, have been well discussed previously.
Although these approximations can significantly
improve the second-order prediction of the NIBA
approach~\cite{Mukamel1987:JCP,Mukamel1989:JCP,Laird1990:JCP,
Cao2000:JCP,JLWu2013:JCP,Reichman1996:JCP},
a systematic resummation approach is still required to
include higher-order corrections and recover the exact quantum dynamics.

\begin{figure}
\includegraphics[width=0.65\columnwidth]{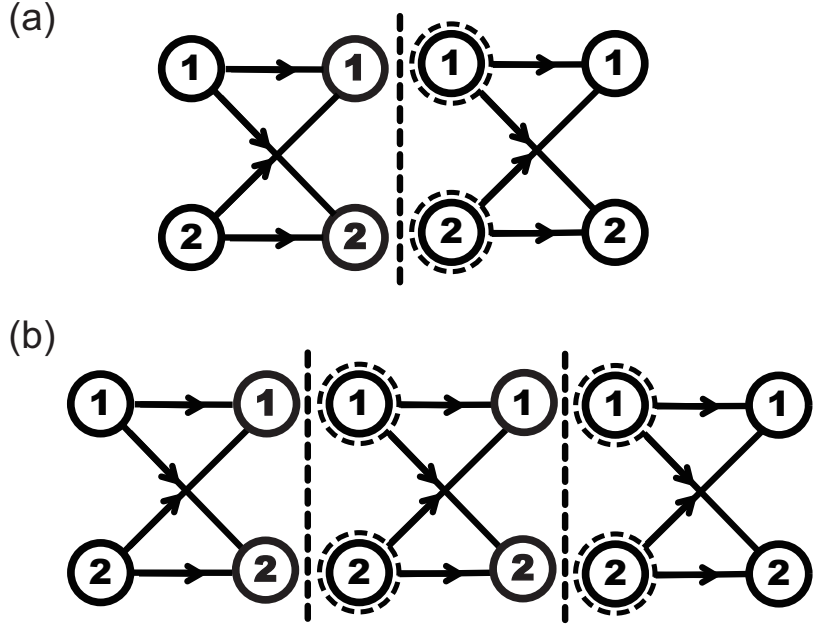}
\caption{The Feynman diagrams of the fourth- (a) and sixth-order (b)
quantum rate kernels in the two-site system (the spin-boson model) under the Pade approximation.
The matrix factorization is realized by inserting vertical dashed lines.
The other symbols are the same as those in Fig.~\ref{fig_001}.}
\label{fig_003}
\end{figure}

We revisit the Pade approximation in Ref.~\cite{Mukamel1987:JCP} to show its physical interpretation,
which will used for a generalized resummation technique.
As mentioned in previous section, the pure dissipation of population, $\delta\calU_\rmP(t)$, vanishes in
Markovian dynamics. For a fast relaxing bath with a weak non-Markovian feature,
or alternatively in the strong dissipation regime where the system transport is slow but Markovian,
an approximate time separation can be expected in the high-order QKE rate kernels. For the leading-order correction
$\calK^{(4)}(\tau_2, \tau_3, \tau_4)$, this approximation is realized mathematically by inserting a
reduced population projection $\calP_\rmP$ before the action of $\delta\calU_\rmP(\tau_3)$~\cite{Mukamel1987:JCP}. In the
reduced population subspace, $\calP_\rmP$ is explicitly written as
\be
\calP_\rmP = \left(\ba{cc}  \rho^\eq_\rmB\}\Tr_\rmB\{ &  0 \\
                               0         &  \rho^\eq_\rmB\}\Tr_\rmB\{ \ea  \right),
\label{eq_013}
\ee
where $\rho^\eq_\rmB\propto \exp(-\beta H_\rmB)$ is the bare bath equilibrium distribution.
Equation~(\ref{eq_013}) results in two identities,
$\calP_\rmP\calP^{(0)}_{\eq}=\calP_\rmP$ and $\calP^{(0)}_{\eq}\calP_\rmP=\calP^{(0)}_{\eq}$.
As a result, the fourth-order QKE rate kernel is factorized into
\be
\calK^{(4)}(\tau_2, \tau_3, \tau_4) \approx  \Xi^{(2)}(\tau_3, \tau_4) \calK^{(2)}(\tau_2),
\label{eq_014}
\ee
with $\Xi^{(2)}(\tau_3, \tau_4)=-\Tr_\rmB\{\calR(\tau_4)\delta\calU_\rmP(\tau_3)\calP_\rmP\}$. The
matrix factorization can be applied to all the higher-order corrections, giving
\be
&&\calK^{(2k)}(\tau_2, \cdots \tau_{2k}) \no \\
&\approx& \Xi^{(2)}(\tau_{2k-1}, \tau_{2k})\cdots\Xi^{(2)}(\tau_3, \tau_4)
\calK^{(2)}(\tau_2).
\label{eq_015}
\ee
Figure~\ref{fig_003} presents the Feynman diagrams of $\calK^{(4)}(\tau_2, \tau_3, \tau_4)$
and $\calK^{(6)}(\tau_2, \cdots \tau_6)$ after
the matrix factorization. With the introduction of the Laplace $z$-transform, the resummation using
the correction term $\tilde{\Xi}^{(2)}(z)$ of  $\tilde{\calK}^{(4)}(z)$ becomes~\cite{Mukamel1987:JCP}
\be
\tilde{\calK}^{(4)}_{\resum}(z) = \left[\calI-\tilde{\Xi}^{(2)}(z)\right]^{-1}\tilde{\calK}^{(2)}(z),
\label{eq_016}
\ee
where $\calI$ is an identity matrix. By expanding Eq.~(\ref{eq_016}) in the $2\times2$ matrix form, we
recover the regular Pade approximation for both forward ($\tilde{k}^{(4)}_{\resum; A\leftarrow D}(z)$)
and backward ($\tilde{k}^{(4)}_{\resum; D\leftarrow A}(z)$) transfer rate kernels.

\begin{figure}
\includegraphics[width=0.75\columnwidth]{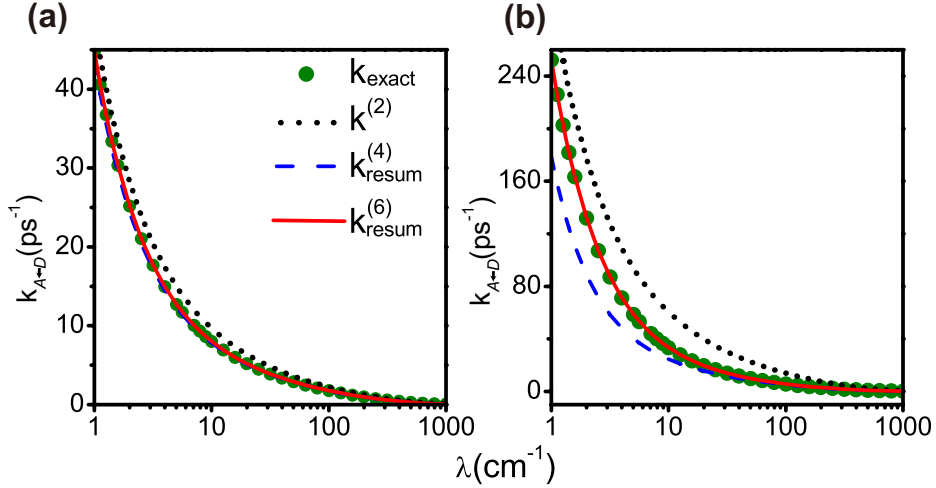}
\caption{The RQKE forward transfer rate from the continued fraction form compared with its exact
value from the hierarchy equation in the unbiased system  with (a) $J=40$ cm$^{-1}$, and (b) $J=100$ cm$^{-1}$.
In each figure, the black dotted line is the second-order Forster rate, the blue dashed line is the
lowest fourth-order RQKE rate (the Pade approximation),
the red solid line is the next sixth-order RQKE rate,
and the green circles are the exact result. The parameters of $\omega_D$ and $T$ are the same as in Fig.~\ref{fig_002}.
}
\label{fig_004a}
\end{figure}

Next we can extend to higher-order corrections with a generalized factorization technique.
Following the definition of $\Xi^{(2)}(\tau_3, \tau_4)$ to higher-orders,
we introduce another expansion series,
\be
&& \Xi^{(2k)}(\tau_3, \cdots, \tau_{2k}) \no \\
&=&(-1)^k\Tr_\rmB\{[\calR(\tau_{2k})\delta\calU_\rmP(\tau_{2k-1})]
 \cdots[\calR(\tau_4)\delta\calU_\rmP(\tau_3)]\calP_\rmP\},
\label{eq_017}
\ee
which is essential for the QKE in the system-bath separated initial condition~\cite{WuRCBS}.
For the sixth-order QKE rate kernel, a more accurate matrix factorization is changed to
$\calK^{(6)}(\tau_2, \cdots, \tau_6)\approx \Xi^{(4)}(\tau_3, \cdots, \tau_{6})\calK^{(2)}(\tau_2)$.
Similar to the cumulant expansion, the `real' fourth-order correction $\Xi^{(4)}$ needs to
exclude the contribution of $\Xi^{(2)}$,
\be
&&\delta\Xi^{(4)}(\tau_3, \cdots, \tau_6) \no \\
&=& \Xi^{(4)}(\tau_3, \cdots, \tau_6)-\Xi^{(2)}(\tau_5, \tau_6)\Xi^{(2)}(\tau_3, \tau_4).
\label{eq_018}
\ee
All the other higher-order QKE rate kernels are subsequently factorized using $\Xi^{(2)}$
and $\delta\Xi^{(4)}$. For conciseness, we introduce the difference of $\delta\Xi^{(4)}$
relative to $\Xi^{(2)}$, which is defined in the Laplace $z$-space as
\be
\delta\tilde{\Xi}^{(4)}(z) = \tilde{\Delta}_4 (z) \tilde{\Xi}^{(2)}(z).
\label{eq_019}
\ee
Here the expansion index 4 is assigned as a subscript since $\tilde{\Delta}_4 (z)$ is
in the same $J$-expansion order as $\tilde{\Delta}_2 (z)=\tilde{\Xi}^{(2)}(z)$.
The approximate full quantum rate kernel resummed from $\tilde{\Delta}_2 (z)$ and
$\tilde{\Delta}_4 (z)$ is derived in a continued fraction form,
\be
\tilde{\calK}^{(6)}_{\resum}(z) =
\left\{\calI-\left[ \calI-\tilde{\Delta}_4(z)\right]^{-1}\tilde{\Delta}_2(z)\right\}^{-1}\tilde{\calK}^{(2)}(z).
\label{eq_020}
\ee
The above factorization scheme can be straightforwardly to an arbitrary expansion order, which
defines the general correction term, $\delta\tilde{\Xi}^{(2k)}(z) = \tilde{\Delta}_{2k-2} (z)\cdots \tilde{\Delta}_2(z)$,
and gives rise to the general matrix continued fraction  form.

The separation of higher-order QKE rate kernels
depicted in Fig.~\ref{fig_003} requires modifications
when the non-Markovian dynamics is not weak.
The dynamic coupling between $\Xi^{(2k-2)}(\tau_3, \cdots, \tau_{2k})$
and $\calK^{(2)}(\tau_2)$ needs to be include, beyond the matrix factorization,
$\calK^{(2k)}(\tau_2, \cdots, \tau_{2k})\approx
\Xi^{(2k-2)}(\tau_3, \cdots, \tau_{2k})\calK^{(2)}(\tau_2)$.
However, this difficulty can be circumvented using the scalar continued fraction form
for each element of the rate kernel. Mathematically, a regular function
can be re-expressed in the continued fraction form, by  matching its Taylor expansion series.
Thus, we propose the scalar continued fraction resummation form for the
forward rate kernel,
\be
\tilde{k}^{(2k)}_{\resum; A\leftarrow D}(z) = \cfrac{1}{1+
\cfrac{\tilde{\delta}_{2; A\leftarrow D}(z)}{\cfrac{\vdots}{1+\tilde{\delta}_{2k-2; A\leftarrow D} (z)}}
}\tilde{k}^{(2)}_{A\leftarrow D}(z),
\label{eq_021}
\ee
where the correction terms are matching the QKE forward rate kernels $\tilde{k}^{(2j)}_{A\leftarrow D}(z)$
term by term, given by
\be
\tilde{\delta}_{2; A\leftarrow D}(z) &=& -\tilde{k}^{(4)}_{A\leftarrow D}(z)/\tilde{k}^{(2)}_{A\leftarrow D}(z), \label{eq_022} \\
\tilde{\delta}_{4; A\leftarrow D}(z) &=& -\tilde{\delta}_{2; A\leftarrow D}(z)-\tilde{k}^{(6)}_{A\leftarrow D}(z)/\tilde{k}^{(4)}_{A\leftarrow D}(z),
\label{eq_023}\\
&\vdots& \no
\ee
The same approach is applied to resum the backward rate kernel $\tilde{k}^{(2k)}_{\resum; D\leftarrow A}(z)$.
Equations~(\ref{eq_021})-(\ref{eq_023}) provides the basic construction of the resummed quantum kinetic expansion
(RQKE) method. To be consistent, the expansion order of the RQKE is defined by
the power of the site-site coupling strength in the highest-order QKE rate kernel considered.
Compared to the generalized Pade approximation in Ref.~\cite{Cho1996:JCP}, the continued fraction
can also be expanded into a rational polynomial form, while the correction terms in the RQKE method
are more straightforwardly obtained without an additional basis expansion.
In addition, as the resummation order $2k$ increases,
all the lower-order correction terms $\tilde{\delta}_{2j (<k-1)}(z)$ are not affected,
which makes the continued fraction form  a more systematic approach.

\begin{figure}
\includegraphics[width=0.75\columnwidth]{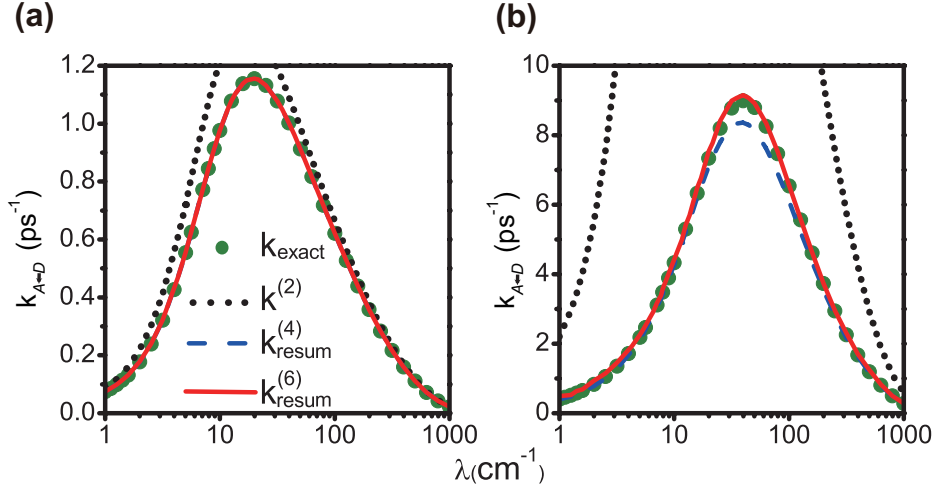}
\caption{The RQKE forward transfer rate from the continued fraction form compared with its exact
value from the hierarchy equation in the biased system, $\varepsilon_{12}=100$ cm$^{-1}$, with
(a) $J=20$ cm$^{-1}$, and (b) $J=100$ cm$^{-1}$.
In each figure, the black dotted line is the second-order Forster rate, the blue dashed line is the
lowest fourth-order RQKE rate (the Pade approximation),
the red solid line is the next sixth-order RQKE rate,
and the green circles are the exact result. The parameters of $\omega_D$ and $T$ are the same as in Fig.~\ref{fig_002}.
}
\label{fig_004b}
\end{figure}

To verify the reliability of the continued fraction form, we use the results of the first three order effective rate
expansions in Section~\ref{sec2} to obtain the RQKE rates $k^{(2k)}_{\resum}=\tilde{k}^{(2k)}_{\resum}(z=0)$,
which are compared with the exact full quantum rates $k_{\exact}$
from the hierarchy equation. In Ref.~\cite{JLWu2013:JCP}, $k_{\exact}$ is calculated under a system-bath separated initial condition,
different from the presumption of the local equilibrium population state in the QKE approach.
The accurate value of $k_{\exact}$ is re-calculated, following the rigorous expression in Ref.~\cite{WuRCBS}.
With the same equilibrium population, the results of $k_{\exact}$ under these two initial conditions are proportional to each other~\cite{WuRCBS}.
The results of $k^{(2)}$, $k^{(4)}_{\resum}$, $k^{(6)}_{\resum}$, and $k_{\exact}$ for the forward transport process
form the donor site 1 to the acceptor site 2 are plotted in Figs.~\ref{fig_004a} and \ref{fig_004b}.
For the unbiased  system ($\varepsilon_{12}=0$), two site-site coupling strengths,
$J=40$ and $100$ cm$^{-1}$ are considered; for the biased system ($\varepsilon_{12}=100$ cm$^{-1}$),
two site-site coupling strengths, $J=20$ and $100$ cm$^{-1}$ are considered.
For the two small site-site coupling strengths, $J=40$ cm$^{-1}$ and $\varepsilon_{12}=0$ in Fig.~\ref{fig_004a}a,
and $J=20$ cm$^{-1}$ and $\varepsilon_{12}=100$ cm$^{-1}$ in Fig.~\ref{fig_004b}a,
the QKE rate kernels converge with the expansion order. The lowest fourth-order RQKE rate
$k^{(4)}_{\resum; A\leftarrow D}$ improves the second-order FGR rate and predict
$k_{\exact; A\leftarrow D}$ accurately in the whole range of the reorganization
energies, $1$ cm$^{-1} \le \lambda \le 1000$ cm$^{-1}$.
For the large coupling strength of $J=100$ cm$^{-1}$ in Figs.~\ref{fig_004a}b and \ref{fig_004b}b,
the QKE rate kernels diverge with the expansion order.
In the unbiased system, $k^{(4)}_{\resum; A\leftarrow D}$ improves $k^{(2)}_{A\leftarrow D}$
mainly in the large-$\lambda$ regime. In the biased system, $k^{(4)}_{\resum; A\leftarrow D}$
largely improves $k^{(2)}_{A\leftarrow D}$, except for a small deviation in the intermediate-$\lambda$ regime.
In Fig. 5b of Ref.~\cite{JLWu2013:JCP}, the difference between $k^{(4)}_{\resum; A\leftarrow D}$ and $k_{\exact; A\leftarrow D}$
in the large-$\lambda$ regime is due to an inconsistent initial condition in the hierarchy equation.
For both cases, the next sixth-order RQKE rate $k^{(6)}_{\resum; A\leftarrow D}$
agrees perfectly with $k_{\exact; A\leftarrow D}$ in the whole $\lambda$ regime.
Our numerical calculations demonstrate that the RQKE rate from the continued fraction form can systematically
approach to the exact value, and the number of necessary correction terms gradually increase with the site-site coupling strength.

\section{Time-Convoluted Quantum Kinetics}
\label{sec4}

The continued fraction form of the bath relaxation effect is verified by the convergence
of the resummed effective rate toward the exact value. In this section, we will further demonstrate
the accuracy of the continued fraction in predicting the detailed time evolution of
site population.

\begin{figure}
\includegraphics[width=0.75\columnwidth]{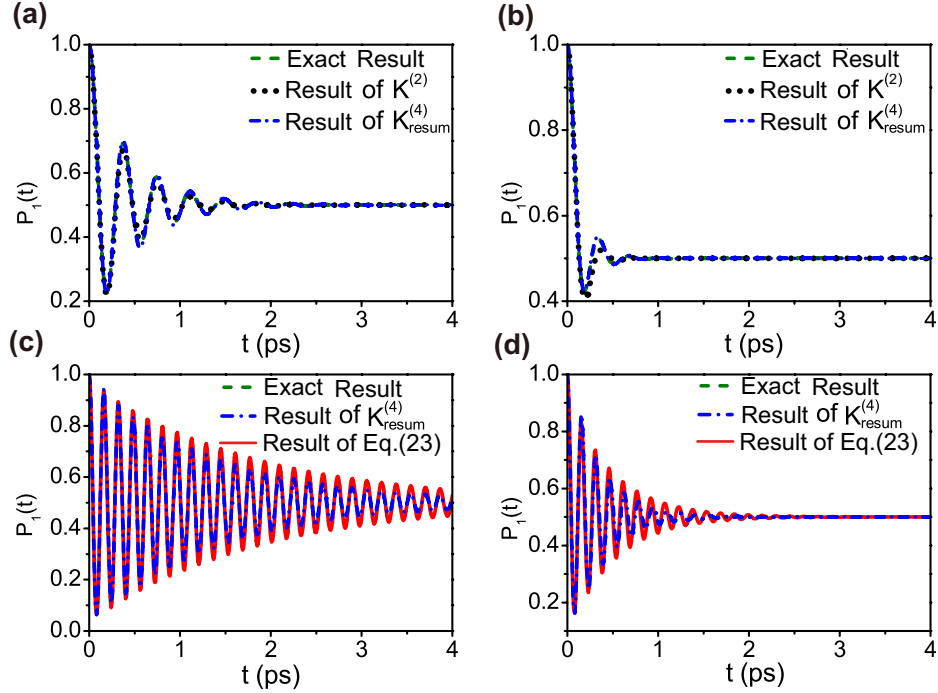}
\caption{The time evolution of the donor population in the unbiased system with
$\varepsilon_{12}=0$. The same quantum Deybe bath in Fig.~\ref{fig_002} is applied.
The conditions of the four figures are (a) $J=20$ cm$^{-1}$ and $\lambda=4$ cm$^{-1}$,
(b) $J=20$ cm$^{-1}$ and $\lambda=12$ cm$^{-1}$, (c) $J=100$ cm$^{-1}$ and $\lambda=4$ cm$^{-1}$,
and (d) $J=100$ cm$^{-1}$ and $\lambda=12$ cm$^{-1}$.
In each figure, the dashed line is from the exact dynamics,
and the dashed-dotted line is from the lowest-order resumed kernel $\tilde{\calK}^{(4)}_{\resum}(z)$ .
In the top two figures, the dotted lines are from the second-order kernel $\tilde{\calK}^{(2)}(z)$.
In the bottom two figures, the solid lines from higher-order resumed rate kernels
fully recover the results of the exact dynamics using Eq.~(\ref{eq_025}) and
coincide with the dashed lines.
}
\label{fig_005}
\end{figure}

All the high-order QKE rate kernels can be derived explicitly, using the cumulant expansion for
the multi-time correlation function of the displacement operator. The time evolution of reduced
site population $P(t)$ is subsequently solved by the convoluted equation in the time $t$-space,
or equivalently by the matrix inversion in the Laplace $z$-space. The computational cost of both
methods is often very high. Instead, we
re-express the QKE rate kernels in a matrix
formalism~\cite{WuRCBS}. The general $2k$-th QKE rate kernel in the Laplace $z$-space is derived in
Ref.~\cite{WuRCBS} as
\be
\tilde{\bm\calK}^{(2k)}(z)
&=& -\bm\calP_\rmP[\tilde{\bm\calR}(z)\delta\tilde{\bm\calU}^{(0)}(z)]^{k-1} \tilde{\bm\calR}(z)\bm\calP^{(0)}_{\eq},
\label{eq_024}
\ee
with $\tilde{\bm\calR}(z)=\bm\calW^{(1)}\delta\tilde{\bm\calU}^{(0)}(z)\bm\calW^{(1)}$.
Here each matrix is defined in an expanded basis set of relevant dynamic variables
and can be mapped to a superoperator
in Section~\ref{sec2}. Specifically, the mapping of two projection matrices are
$\calP_\rmP\Leftrightarrow \bm\calP_\rmP $ and $\calP^{(0)}_{\eq}\Leftrightarrow \bm\calP^{(0)}_{\eq}$.
The two interaction Liouville superoperators are combined together and mapped to a perturbed transition rate matrix,
 $\{i\calL_{\rmP\rmC}, i\calL_{\rmC\rmP} \}\Leftrightarrow \bm\calW^{(1)}$. The two
unperturbed time propagation superoperaotrs
are also combined together and mapped to an unperturbed pure dissipative matrix,
$\{\delta\tilde{\calU}_\rmP(z), \tilde{\calU}_\rmC(z)\} \Leftrightarrow \delta\tilde{\bm\calU}^{(0)}(z)$.

For over-damped quantum dynamics in the two-site system, the time evolution of site population
is close to a single exponential decaying function (Markovian behavior),
which can be described by the time-integrated effective rate.
To illustrate the relevant non-Markovian behavior, we focus on small and intermediate reorganization energies with
under-damped dynamics. In our two-site system, we choose two typical reorganization energies,
$\lambda=4$ and $12$ cm$^{-1}$, for each system condition ($\varepsilon_{12}$ and $J$) in Figs.~\ref{fig_004a}
and \ref{fig_004b}. The exact time evolution of site population, $P_{\exact; 1}(t)$,
is solved by the hierarchy equation using the local equilibrium
population state initially at the donor site 1. Next we re-calculate the site population $\tilde{P}_{\exact; 1}(z)$
in the Laplace $z$-space, and obtain a new estimation of the time evolution,
$P^{\prime}_{\exact; 1}(t)=\mathrm{LT}^{-1}[\tilde{P}_{\exact; 1}(z)]$, using the inverse
Laplace transform, $\mathrm{LT}^{-1}[\cdots]$.
The two time evolution predictions, $P_{\exact; 1}(t)$ and $P^{\prime}_{\exact; 1}(t)$
are found to be identical, confirming the reliability of the numerical
inverse Laplace transform. In our model system, Eq.~(\ref{eq_024}) is also numerically
solved using the hierarchy equation approach~\cite{WuRCBS}.
The estimation of the site population from the $2k$-th order RQKE rate kernel is written as
\be
P^{(2k)}_{\resum; 1}(t)=\mathrm{LT}^{-1}\left[\frac{z+\tilde{k}^{(2k)}_{\resum; A\leftarrow D}(z)}
{z[z+\tilde{k}^{(2k)}_{\resum; A\leftarrow D}(z)+\tilde{k}^{(2k)}_{\resum; D\leftarrow A}(z)]} \right].
\label{eq_025}
\ee

\begin{figure}
\includegraphics[width=0.75\columnwidth]{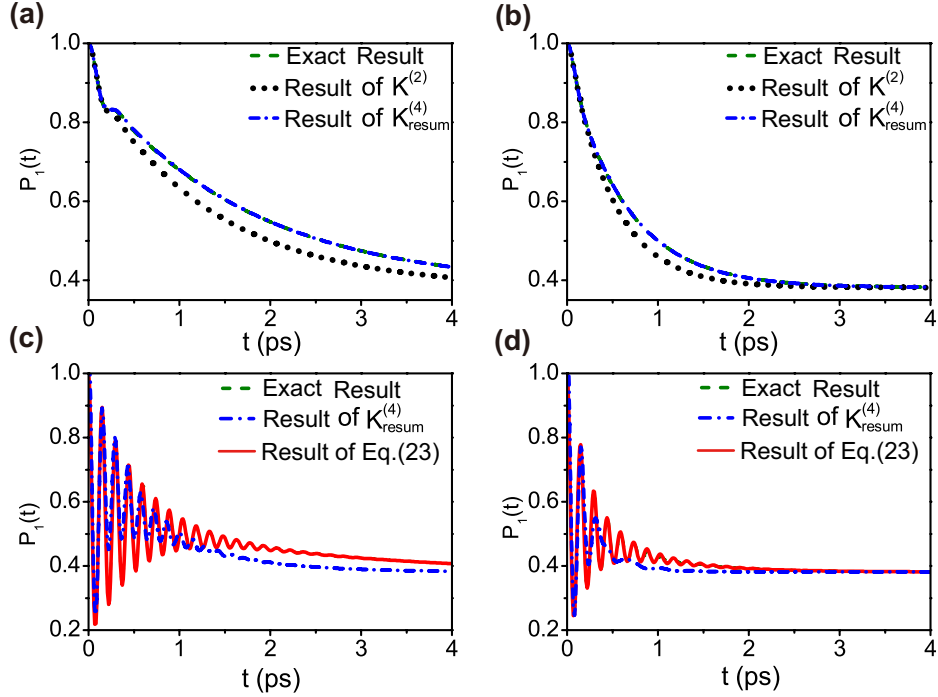}
\caption{The time evolution of the donor population in the biased system with
$\varepsilon_{12}=100$ cm$^{-1}$. The same quantum Deybe bath in Fig.~\ref{fig_002} is applied.
The conditions of the four figures are (a) $J=20$ cm$^{-1}$ and $\lambda=4$ cm$^{-1}$,
(b) $J=20$ cm$^{-1}$ and $\lambda=12$ cm$^{-1}$, (c) $J=100$ cm$^{-1}$ and $\lambda=4$ cm$^{-1}$,
and (d) $J=100$ cm$^{-1}$ and $\lambda=12$ cm$^{-1}$.
 In each figure, the dashed line is from the exact dynamics,
and the dashed-dotted line is from the lowest-order resumed kernel $\tilde{\calK}^{(4)}_{\resum}(z)$ .
In the top two figures, the dotted lines are from the second-order kernel $\tilde{\calK}^{(2)}(z)$.
In the bottom two figures, the solid lines from higher-order RQKE rate kernels fully recover the results of
the exact dynamics using Eq.~(\ref{eq_025}) and coincide with the dashed lines. }
\label{fig_006}
\end{figure}

We apply the same two-site system with the same quantum Debye bath with $\omega^{-1}_D=100$ fs and $T=300$ K in previous two sections.
The comparison between $P^{(2)}_1(t)$, $P^{(2k)}_{\resum; 1}(t)$ and  $P_{\exact; 1}(t)$ is organized in Figs.~\ref{fig_005}
and \ref{fig_006}, where $P^{(2)}_1(t)$ is the second-order NIBA prediction.
In the unbiased system ($\varepsilon=0$) with the small site-site coupling ($J=40$ cm$^{-1}$),
$P^{(2)}_1(t)$ is close to the exact time evolution $P_{\exact; 1}(t)$ with a small deviation.
The lowest fourth-order RQKE rate kernel, $\tilde{\calK}^{(4)}_{\resum}(z)$, further improves $P^{(2)}_1(t)$
and provides almost identical results of $P_{\exact; 1}(t)$ for the two reorganization energies.
As the site-site coupling is increased to $J=100$ cm$^{-1}$,
$P^{(4)}_{\resum; 1}(t)$ improved from the NIBA prediction
also deviates from the exact result
$P_{\exact; 1}(t)$.
We find that $P^{(2k)}_{\resum; 1}(t)$ gradually approaches to $P_{\exact; 1}(t)$
as the $J$-expansion order $2k$ increases in the continued fraction form.
As shown in Figs.~\ref{fig_005}c and d,
 $P^{(8)}_{\resum; 1}(t)$ and $P^{(6)}_{\resum; 1}(t)$ from the eighth- and sixth-order
RQKE rate kernels fully recover $P_{\exact; 1}(t)$ for $\lambda=4$ and $12$ cm$^{-1}$,
respectively. In the biased system ($\varepsilon=100$ cm$^{-1}$)
with the small site-site coupling ($J=20$ cm$^{-1}$), $P^{(2)}_1(t)$ clearly deviates from
$P_{\exact; 1}(t)$, while $P^{(4)}_{\resum; 1}(t)$ from the Pade approximation
becomes almost identical to $P_{\exact; 1}(t)$  for the two values of $\lambda$.
Although the time-integrated rate $k^{(4)}_{\resum}$ is very close to the exact value $k_{\exact}$
in the small-$\lambda$ regime, the prediction of $P^{(4)}_{\resum; 1}(t)$ is no
longer reliable for the strong site-site coupling ($J=100$ cm$^{-1}$).
Similarly, we extend the continued fraction form to higher orders, and $P^{(10)}_{\resum; 1}(t)$
from the tenth-order RQKE rate kernel fully recovers $P_{\exact; 1}(t)$
for $\lambda=4$ and $12$ cm$^{-1}$.
Thus, the exact quantum dynamics can be fully predicted by the RQKE rate kernels
in the continued fraction form. The convergence order of the continued fraction
for the detailed time evolution in general increases as the reorganization energy decreases.
Since the equilibrium population in the unbiased system is unchanged
with the system and bath parameters, the convergence order is usually smaller than
that in the biased system.

\section{Temperature Dependence of the Quantum Equilibrium Population}
\label{sec5}

In this section, we will further demonstrate the accuracy of the continued fraction
in predicting the temperature dependence of quantum equilibrium population.

In the original matrix continued fraction form, the expansion $\tilde{\Xi}^{(2k)}(z)$
from the factorization scheme on the high-order
QKE rate kernels leads to the same correction terms for both forward and backward
transfer rate kernels, i.e., $\tilde{\delta}_{2k; A\leftarrow D} (z)=\tilde{\delta}_{2k; D\leftarrow A} (z)$.
The ratio of the two time-integrated RQKE rates,
$k^{(2k)}_{\resum; A\leftarrow D}/k^{(2k)}_{\resum; D\leftarrow A}$, is unchanged as the
resummation order increases. The equilibrium population is always
the same as the classical Boltzmann distribution of the FGR prediction,
$P_{\eq; n}\propto \exp(-\beta \varepsilon_{n})$, which is only valid at high temperatures.
In our modified scalar continued fraction form, the correction terms of
the forward and backward rate kernels are determined independently, which allows
$\tilde{\delta}_{2k; A\leftarrow D} (z) \neq \tilde{\delta}_{2k; D\leftarrow A} (z)$.
Consequently, the equilibrium population predicted by the RQKE rate can deviate from
the classical Boltzmann distribution and approach to the exact quantum Boltzmann
distribution, $P_{\eq; n}\propto [\Tr_B\{\exp(-\beta H_{\tot})\}]_{nn}$~\cite{CKLee2012:JCP, Jeremy2012:PRB}.

\begin{figure}
\includegraphics[width=0.65\columnwidth]{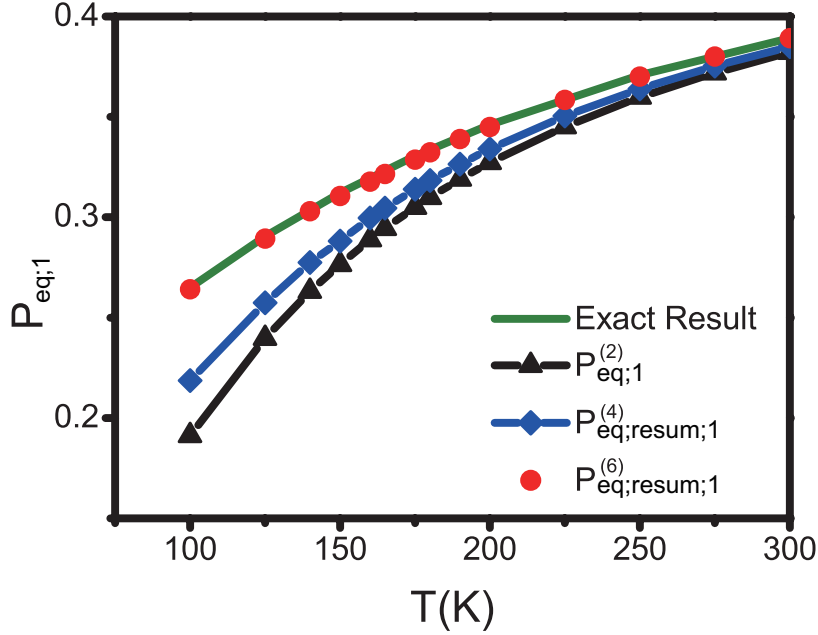}
\caption{The equilibrium donor population versus the temperature. The solid line with up-triangle symbols
is the result of Fermi's golden rule rate. The solid line with diamond symbols is the result of the
lowest fourth-order RQKE rates using the Pade approximation. The circle symbols
represent the result of the next sixth-order RQKE rates.
The solid line without symbols is the exact result from the hierarchy equation. The parameters
are $\varepsilon_{12}=100$ cm$^{-1}$, $J=100$ cm$^{-1}$,
$\lambda=100$ cm$^{-1}$, and $\omega^{-1}_D=100$ fs.}
\label{fig_007}
\end{figure}

As a verification, we extend our previous study at a high temperature $T=300$ K to
lower temperatures.
Since the equilibrium population is always one half in the unbiased system, we
only consider the biased system, $\varepsilon_{12}=100$ cm$^{-1}$ with
$J=100$ cm$^{-1}$ and $\lambda=100$ cm$^{-1}$.
The $2k$-th order prediction of the donor equilibrium population $P^{(2k)}_{\eq; \resum; 1}$
is obtained using the time-integrated RQKE rates,
\be
P^{(2k)}_{\eq;\resum; 1} = \frac{k^{(2k)}_{\resum; D\leftarrow A}}{k^{(2k)}_{\resum; A\leftarrow D}
+k^{(2k)}_{\resum; D\leftarrow A}}.
\ee
The full expression of the time correlation function $g(t)$ is applied in the calculation
of the QKE rate kernels, without the high-temperature approximation.
Similarly, $10^{12}$ Monte Carlo samples are simulated for an accurate estimation of $k^{(6)}_{\resum}$.
The hierarchy equation with the Matsubara frequency summation is used to obtain the
exact equilibrium population, which is numerically the same as the result of the
stochastic path integral~\cite{CKLee2012:JCP,Jeremy2012:PRB}. Our numerical calculation shows that
each correction term $\delta_{2j(<k-1)}$ is different for the forward and backward rates,
and the deviation increases as temperature decreases.
As shown in Fig.~\ref{fig_007}, the RQKE rates systematically
improves the prediction of $P^{(2k)}_{\eq; \resum; 1}$ from the second-order FGR result
to the exact result. With specific parameters in our calculation,
the sixth-order RQKE rates provide an excellent prediction of the
exact result over the whole temperature range (100 K $\le T\le 300$ K).
With more correction terms included, we expect that the scalar continued fraction
resummation can be straightforwardly extended to lower temperatures.

\section{Summary}
\label{sec6}

In this paper, we extend our previous study of the quantum kinetic expansion (QKE) approach in
the two-site system (the spin-boson model).
The factorization scheme for the high-order QKE rate kernels
in the weak non-Markovian dynamics
leads to the matrix continued fraction form for the resummation technique of QKE rate kernels.
To be valid in an arbitrary condition, we further introduce  the scalar continued fraction form
for forward and backward rate kernels separately,
where the correction terms are obtained by matching the higher-order QKE rate kernels.
Consequently, a systematic resummed quantum kinetic expansion (RQKE) method is constructed,
and the expansion order of the RQKE method is consistent with the highest order
of the QKE rate kernel. To the lowest fourth-order, the continued
fraction form recovers the Pade approximation, while the higher-order RQKE
correction terms represent the additional bath relaxation effects.
As shown by numerical calculations
in this paper, the prediction of the RQKE method systematically
improves with the expansion order and can fully reproduce the exact quantum dynamics calculated
from the hierarchy equation. With specific parameters considered in this paper,
the time-integrated RQKE rate at the sixth order
can be almost identical to the exact result for both unbiased
and biased system, with both weak and strong site-site coupling
strengths. More importantly, the detailed time evolution can be exactly predicted as well,
as higher-order correction terms are gradually included. The temperature dependence of
the equilibrium population is also verified, as the classical Boltzmann distribution of
the second-order FGR prediction is improved toward the exact quantum Boltzmann distribution.
The convergence order
generally increases  with the increase of the site-site coupling strength, the
decrease of the reorganization energy and the decrease of temperature.

The numerical calculations of this paper are focused on the harmonic bath
with a quantum Debye spectral density. The formal expression of the QKE rate kernel in
Eq.~(\ref{eq_006})  is however invariant of the bath structure, whether Gaussian or non-Gaussian,
so that the RQKE method can be applied to a general bath, combined with other numerical methods.
The mathematical strategy of applying the continued fraction form is not limited to the
two-site system, and its application to more complicated systems will be demonstrated in our forthcoming papers.
The RQKE method provides a systematically converged approach of quantum dynamics,
and its continued fraction form can inspire possibilities of
other higher-order resummation techniques, such as the extension of the
Landau-Zener approximation and modifications originally for the lowest order correction.

\begin{acknowledgements}

The work reported here is supported by the Ministry of Science and Technology of China (MOST-2014CB921203),
the National Science Foundation of China (NSFC-21173185),
and Research Fund for the Doctoral Program of Higher Education of China (J20120102).

\end{acknowledgements}

\appendix
\section{Fourth- and Sixth-Order Quantum Rate Kernels in the two-Site System}
\label{appa}

In this appendix, we summarize the expressions of the fourth- and sixth-order QKE rate kernels in the two-site system
with a $\delta$-spatial correlation. Notice that such a two-site system coupled with the harmonic bath
is equivalent to the standard spin-boson model with a doubled reorganization energy.
The fourth-order QKE rate kernel for a general multi-site system is derived in
Ref.~\cite{JLWu2013:JCP}, and we simplify this expression with the consideration of the two-site system.
The forward transfer rate kernel from the donor site 1 to the acceptor site 2 is written explicitly as
\be
&& \mathcal{K}_{21}^{(4)}(\tau_2, \tau_3, \tau_4)/2|J|^4 \no \\
&=&\mathrm{Re}\{e^{i\tilde{\varepsilon}_{12}\tau^-_{4}-2\calG^+_4}
   (e^{2F_{4}^{-}}-1) +e^{i\tilde{\varepsilon}_{12}\tau^-_{4}-2\calG^-_4}
   (e^{2F_{4}^{+}}-1) \no \\
& &~~+e^{i\tilde{\varepsilon}_{12}\tau^+_{4}-2\calG^+_4}
   (e^{-2F_{4}^{+}}-1)+e^{i\tilde{\varepsilon}_{12}\tau^+_{4}-2\calG^-_4}
   (e^{-2F_{4}^{-}}-1)\},
\label{eq_app101}
\ee
with $\tau^{\pm}_{4}=\tau_2\pm\tau_4$, $\calG_4^{\pm}= g(\tau_{2})+g(\pm\tau_{4})$, and
$F_{4}^{\pm}=g(\pm\tau_{3})-g(\tau_{2}+\tau_{3})-g(\pm(\tau_{3}+\tau_{4}))+g(\tau_{2}+\tau_{3}+\tau_{4})$.

The sixth-order quantum rate kernel after expanding each term is given by
\be
&&\calK^{(6)}(\tau_2,\cdots, \tau_6) \no \\
&=& \Tr_B\{\calR(\tau_{6})\mathcal{U}_{P}(\tau_{5})\calR(\tau_{4})\mathcal{U}_{P}(\tau_{3})
\calR(\tau_{2})\calP^{(0)}_{\eq}\} \no \\
&&- \mathcal{K}^{(2)}(\tau_{6})\mathcal{K}^{(2)}(\tau_{4})\mathcal{K}^{(2)}(\tau_{2}) \no \\
&&+\mathcal{K}^{(2)}(\tau_{6})\mathcal{K}^{(2)}(\tau_{4},\tau_{3},\tau_{2}) \no \\
&&+\mathcal{K}^{(4)}(\tau_{6},\tau_{5},\tau_{4})\mathcal{K}^{(2)}(\tau_{2}).
\label{eq_app102}
\ee
For conciseness, we only present one off-diagonal element of
\be
\calY &=& \Tr_B\{\calR(\tau_{6})\mathcal{U}_{P}(\tau_{5})\calR(\tau_{4})\mathcal{U}_{P}(\tau_{3})
\calR(\tau_{2})\calP^{(0)}_{\eq}\} \no \\
&&- \mathcal{K}^{(2)}(\tau_{6})\mathcal{K}^{(2)}(\tau_{4})\mathcal{K}^{(2)}(\tau_{2}),
\ee
and all the other terms can be found from the second- and fourth-order QKE rate kernels.
For the quantum transport process from the donor site 1 to the acceptor site 2, the corresponding term $\calY_{21}$
is explicitly given by

\be
& &\calY_{21}/(-2|J|^6)  \no \\
&= &\mathrm{Re}\left\{ e^{i\tilde{\epsilon}_{12}\tau^{++}_{6} -2\calG^{++}_6} (e^{-F_{6A}^{+;+}}-1)
+ e^{i\tilde{\epsilon}_{12}\tau^{-+}_{6} -2\calG^{++}_6} (e^{-F_{6D}^{+;-}}-1) \right. \no \\
&&~~~~ + e^{i\tilde{\epsilon}_{12}\tau^{+-}_{6} -2\calG^{++}_6} (e^{-F_{6A}^{-;-}}-1)
+ e^{i\tilde{\epsilon}_{12}\tau^{--}_{6} -2\calG^{++}_6} (e^{-F_{6D}^{-;+}}-1)  \no \\
&&~~~~+ e^{i\tilde{\epsilon}_{12}\tau^{+-}_{6} -2\calG^{--}_6} (e^{-F_{6B}^{-;+}}-1)
+ e^{i\tilde{\epsilon}_{12}\tau^{--}_{6} -2\calG^{--}_6} (e^{-F_{6C}^{-;-}}-1)   \no \\
&&~~~~+ e^{i\tilde{\epsilon}_{12}\tau^{++}_{6} -2\calG^{--}_6} (e^{-F_{6B}^{+;-}}-1)
+ e^{i\tilde{\epsilon}_{12}\tau^{-+}_{6} -2\calG^{--}_6} (e^{-F_{6C}^{+;+}}-1)   \no \\
&&~~~~+ e^{i\tilde{\epsilon}_{12}\tau^{+-}_{6} -2\calG^{+-}_6} (e^{-F_{6A}^{-;+}}-1)
+ e^{i\tilde{\epsilon}_{12}\tau^{--}_{6} -2\calG^{+-}_6} (e^{-F_{6D}^{-;-}}-1)   \no \\
&&~~~~+ e^{i\tilde{\epsilon}_{12}\tau^{++}_{6} -2\calG^{+-}_6} (e^{-F_{6A}^{+;-}}-1)
+ e^{i\tilde{\epsilon}_{12}\tau^{-+}_{6} -2\calG^{+-}_6} (e^{-F_{6D}^{+;+}}-1)   \no \\
&&~~~~+ e^{i\tilde{\epsilon}_{12}\tau^{++}_{6} -2\calG^{-+}_6} (e^{-F_{6B}^{+;+}}-1)
+ e^{i\tilde{\epsilon}_{12}\tau^{-+}_{6} -2\calG^{-+}_6} (e^{-F_{6C}^{+;-}}-1) \no \\
&&~~~~\left.+ e^{i\tilde{\epsilon}_{12}\tau^{+-}_{6} -2\calG^{-+}_6} (e^{-F_{6B}^{-;-}}-1)
+ e^{i\tilde{\epsilon}_{12}\tau^{--}_{6} -2\calG^{-+}_6} (e^{-F_{6C}^{-;+}}-1)\right\}.
\ee
Here we introduce the abbreviated notations, $\tau^{\pm\pm}_{6} = \tau_{2}\pm\tau_{4}\pm\tau_6$, and
$\calG^{\pm\pm}_6 =g(\tau_{2})+g(\pm\tau_{4})+g(\pm\tau_{6})$, where the left and right $\pm$ superscript
symbols are associated with $\tau_4$ and $\tau_6$, respectively. Additional abbreviated notations,
$\tau_{ij}=\tau_i+\tau_j$,  $\tau_{ijk}=\tau_i+\tau_j+\tau_k$, $\cdots$ ($i, j, k = 2, 3, \cdots, 6$),
are introduced to express the functions of $F_6$ as
\begin{subequations}
  \be
F_{6A}^{\pm;\pm} &=& 2g(\tau_{3}) \pm 2g(\pm \tau_{5}) -2g(\tau_{23})-2g(\tau_{34})  \no \\
 && \mp 2g(\tau_{45}) \mp 2g( \pm \tau_{56})+2g(\tau_{234})   \no \\
 && \pm 2g(\tau_{345}) \pm 2g(\tau_{456}) \mp 2g(\tau_{2345})  \no \\
 && \mp 2g(\tau_{3456}) \pm 2g(\tau_{23456}),
\ee
\be
F_{6B}^{\pm;\pm} &=& 2g(-\tau_{3}) \pm 2g(\pm \tau_{5}) -2g(\tau_{23})-2g(-\tau_{34})  \no \\
 && \mp 2g(- \tau_{45}) \mp 2g( \pm \tau_{56})+2g(\tau_{234})   \no \\
 && \pm 2g(-\tau_{345}) \pm 2g(-\tau_{456}) \mp 2g(\tau_{2345})  \no \\
 && \mp 2g(-\tau_{3456}) \pm 2g(\tau_{23456}),
\ee
\be
F_{6C}^{\pm;\pm} &=& -2g(\tau_{3}) \mp 2g(\pm \tau_{5}) + 2g(\tau_{23}) + 2g(\tau_{34})  \no \\
 && \pm 2g(- \tau_{45}) \pm 2g( \pm \tau_{56}) - 2g(\tau_{234})   \no \\
 && \pm 2g(\tau_{345}) \mp 2g(-\tau_{456}) \mp 2g(\tau_{2345})  \no \\
 && \mp 2g(\tau_{3456}) \pm 2g(\tau_{23456}),
\ee
\be
F_{6D}^{\pm;\pm} &=& -2g(-\tau_{3}) \mp 2g(\pm \tau_{5}) + 2g(\tau_{23}) + 2g(-\tau_{34})  \no \\
 && \pm 2g(\tau_{45}) \pm 2g( \pm \tau_{56}) - 2g(\tau_{234})   \no \\
 && \pm 2g(-\tau_{345}) \mp 2g(\tau_{456}) \mp 2g(\tau_{2345})  \no \\
 && \mp 2g(-\tau_{3456}) \pm 2g(\tau_{23456}),
\ee
\end{subequations}
where the left $\pm$ superscript symbol is associated with operations between $g$ functions, and
the right $\pm$ superscript symbol is associated with the sign of the time variable inside $g$ functions.


\begin{thebibliography}{56}
\expandafter\ifx\csname natexlab\endcsname\relax\def\natexlab#1{#1}\fi
\expandafter\ifx\csname bibnamefont\endcsname\relax
  \def\bibnamefont#1{#1}\fi
\expandafter\ifx\csname bibfnamefont\endcsname\relax
  \def\bibfnamefont#1{#1}\fi
\expandafter\ifx\csname citenamefont\endcsname\relax
  \def\citenamefont#1{#1}\fi
\expandafter\ifx\csname url\endcsname\relax
  \def\url#1{\texttt{#1}}\fi
\expandafter\ifx\csname urlprefix\endcsname\relax\def\urlprefix{URL }\fi
\providecommand{\bibinfo}[2]{#2}
\providecommand{\eprint}[2][]{\url{#2}}


\bibitem{sachdev:quantum_phase_transition_book}
S. Sachdev, {\it Quantum Phase Transitions} (Cambridge University Press, New York, 2011).

\bibitem{Leggett1981:PhysRevLett}
A. O. Caldeira and A. J. Leggett, Phys. Rev. Lett. {\bf 46}, 211 (1981).

\bibitem{Leggett1987:RMP}
A. J. Leggett, S. Chakravarty, A. T. Dorsey, M. P. A. Fisher,
A. Garg, and W. Zwerger, Rev. Mod. Phys. {\bf 59}, 1 (1987).

\bibitem{Breuer2002}
H. P. Breuer and F. Petruccione, {\it The Theory of Open Quantum Systems}
(Oxford University Press, New York, 2002).

\bibitem{Nitzan2006}
 A. Nitzan, {\it Chemical Dynamics in Condensed Phases: Relaxation, Transfer and
 Reactions in Condensed Molecular Systems} (Oxford University Press, New York, 2006).


\bibitem{Forster1948:Annphys}
T. F\"{o}rster, Ann. Phys. (Leipzig) {\bf 437}, 55 (1948).



 \bibitem{Marcus1964:ARPC}
R. A. Marcus, Annu. Rev. Phys. Chem. {\bf 15}, 155 (1964).






\bibitem{Zusman1979:CP}
L. D. Zusman, Chem. Phys. {\bf 49}, 295 (1980).

\bibitem{Sumi1985:JCP}
H. Sumi and R. A. Marcus, J. Chem. Phys. {\bf 84}, 4894 (1986).


\bibitem{CaoJung:JCP}
J. S. Cao and Y. Jung, J. Chem. Phys. {\bf 112}, 4716 (2000).


\bibitem{Loring1987: JCP}
R. F. Loring and S. Mukamel, J. Chem. Phys. {\bf 87}, 1272 (1987).


\bibitem{Mukamel1987:JCP}
M. Sparpaglione and S. Mukamel, J. Chem. Phys. {\bf 88}, 3263 (1988).

\bibitem{Mukamel1989:JCP}
Y. Hu and S. Mukamel, J. Chem. Phys. { \bf 91 }, 6973 (1989).

\bibitem{Laird1990:JCP}
B. B. Laird, J. Budimir, and J. L. Skinner, J. Chem. Phys. {\bf 94}, 4391 (1991).

\bibitem{Cao2000:JCP}
J. S. Cao, J. Chem. Phys. {\bf 112}, 6719 (2000).

\bibitem{JLWu2013:JCP}
J. L. Wu and J. S. Cao, J. Chem. Phys. {\bf 139}, 044102 (2013).


\bibitem{Reichman1996:JCP}
D. R. Reichman and R. J. Silbey, J. Chem. Phys. {\bf 104}, 1506 (1996)



\bibitem{Cao2009:JPCA}
J. S. Cao and R. J. Silbey, J. Phys. Chem. A {\bf 113}, 13825 (2009).


\bibitem{JLWu2010:NJP}
J. L. Wu, F. Liu, Y. Shen, J. S. Cao, and R. J. Silbey,
New J. Phys. {\bf 12}, 105012 (2010).


\bibitem{Moix2011}
J. Moix, J. L. Wu, P. F. Huo, D. Coker, and J. S. Cao,
J. Phys. Chem. Lett. {\bf 2}, 3045 (2011).


\bibitem{JLWu2012:JCP}
J. L. Wu, F. Liu, J. Ma, R. J. Silbey, and J. S. Cao,
J. Chem. Phys. {\bf 137}, 174111 (2012).


\bibitem{JLWu2013:PRL}
 J. L. Wu, R. J. Silbey, and J. S. Cao,
 Phys. Rev. Lett. {\bf 110}, 200402 (2013).



\bibitem{Plenio2008:NJP}
M. B. Plenio and S. F. Huelga, New J. Phys. {\bf 10}, 113019 (2008).

\bibitem{Rebentrost2009:NJP}
P. Rebentrost, M. Mohseni, I. Kassal, S. Lloyd, and A. Aspuru-Guzik, New J. Phys.
{\bf 11}, 033003 (2009).



\bibitem{May2004}
V. May and K. Oliver, {\it Charge and Energy Transfer Dynamics in Molecular Systems}
(Wiley-VCH, Weinheim, 2004).



\bibitem{Feynman1963:AnnPhys}
R. P. Feynman and F. L. Vernon, Jr., Ann. Phys. (N.Y.)
{\bf 24}, 118 (1963).

\bibitem{Nakajima1958}
S. Nakajima, Prog. Theor. Phys. {\bf 20}, 948 (1958).

 \bibitem{Zwanzig1960:JCP}
 R. Zwanzig, J. Chem. Phys. {\bf 33}, 1338 (1960).

\bibitem{Redfield1957}
A. G. Redfield, IBM J. Res. Dev. {\bf 19}, 1 (1957).


\bibitem{Harris1984:JCP}
R. J. Silbey and R. A. Harris, J. Chem. Phys. {\bf 80}, 2615 (1984).


\bibitem{Wang2014}
C. Wang, J. Ren, and J. S. Cao, arXiv:1410.4366 (2014)



\bibitem{Sumi1980:JPSP}
H. Sumi, J. Phys. Soc. Jpn. {\bf 49}, 1701 (1980)

\bibitem{Troy2014:JCP}
M. G. Mavros and T. V. Voorhis, J. Chem. Phys. {\bf 141}, 054112 (2014).


\bibitem{Cho1996:JCP}
M. Cho and R. J. Silbey, J. Chem. Phys. {\bf 106}, 2654 (1997).


\bibitem{Tanimura1989:JPSJ}
Y. Tanimura and R. Kubo, J. Phys. Soc. Jpn. {\bf 58}, 101 (1989).


\bibitem{Shao2004:CPL}
Y. Yan, F. Yang, Y. Liu, and J. Shao, Chem. Phys. Lett. {\bf 395}, 216 (2004).

\bibitem{Yan2005:JCP}
R. X. Xu, P. Cui, X. Q. Li, Y. Mo, and Y. J. Yan, J. Chem. Phys. {\bf 122}, 041103 (2005).


\bibitem{Moix2013:JCP}
J. M. Moix and J. Cao, J. Chem. Phys.  {\bf 139}, 134106 (2013).



\bibitem{CKLee2012:JCP}
C. K. Lee, J. Moix and J. S. Cao, J. Chem. Phys. {\bf 136}, 204120 (2012).

\bibitem{Jeremy2012:PRB}
J. M. Moix, Y. Zhao, and J. S. Cao, Phys. Rev. B {\bf 85}, 115412 (2012).


\bibitem{WuRCBS}
J. L. Wu, in preparation.

\end{thebibliography}
\end{document}